\documentclass[12pt]{iopart}
\pdfoutput=1
\usepackage{graphicx}

\begin{document}

\title[Entanglement Thermodynamics]{Entanglement Thermodynamics}

\author{John Schliemann}

\address{Institute for Theoretical Physics, University of Regensburg,
D-93040 Regensburg, Germany}
\ead{john.schliemann@physik.uni-regensburg.de}
\begin{abstract}
We investigate further the relationship between the 
entanglement spectrum of a composite many-body system and the
energy spectrum of a subsystem making use of concepts of canonical
thermodynamics.
In many important cases the entanglement Hamiltonian is, in the limit of
strong coupling between subsystems, proportional to the energy Hamiltonian
of the subsystem. The proportionality factor is an appropriately defined
coupling parameter, suggesting to interpret the latter as a inverse
temperature. We identify a condition on the entanglement Hamiltonian which
rigorously guarantees this interpretation to hold and removes any
ambiguity in the definition of the entanglement Hamiltonian regarding  
contributions proportional to the unit operator. 
Illustrations of our findings are provided by spin ladders
of arbitrary spin length, and by 
bilayer quantum Hall systems at total filling factor $\nu=2$. 
Within mean-field description, the latter system realizes an entanglement
spectrum of free fermions with just two levels of equal modulus
where the analogies to canonical thermodynamics are particularly close. 
\end{abstract}

\section{Introduction}
\label{intro}

Quantum entanglement is by now an established ingredient to the understanding
and description of various phenomena in condensed matter and many body physics
\cite{Amico08,Tichy11,Eisert10}. A more recent development represents the
notion of the entanglement spectrum, i.e. the spectrum of the reduced density
matrix obtained from the ground state of a composite system upon tracing out
a subsystem \cite{Li08}. Moreover, since this reduced density matrix does not 
have any negative eigenvalues, it can always be formulated as
\begin{equation}
\rho_{\rm red}=\frac{e^{-{\cal H}_{\rm ent}}}{Z}
\label{rhoentham}
\end{equation}
with a partition function $Z={\rm tr}(e^{-{\cal H}_{\rm ent}})$ and 
an entanglement Hamiltonian ${\cal H}_{\rm ent}$. The physical significance of the
latter stems from the observation that
in many important cases, the entanglement spectrum shows, in
the regime of strong coupling of the constituent subsystems, a striking
similarity to the energy spectrum of the subsystem itself or its complement
\cite{Poilblanc10,Cirac11,Peschel11,Lauchli11,Schliemann12,Schliemann11,Schliemann13},
\begin{equation}
{\cal H}_{\rm ent}\approx\lambda{\cal H}_1/ t\,.
\label{enterg}
\end{equation}
Here ${\cal H}_1$ is the energetic Hamiltonian of the subsystem and
$t$ is a typical energy scale of it. 
The dimensionless parameter
$\lambda$ describes the coupling between the subsystems and is small
in the limit of strong coupling, and Eqs.~(\ref{rhoentham}), (\ref{enterg})
clearly suggest to interpret it as an inverse temperature $\beta$.

As a typical example , in the case of
spin ladders, $\lambda$ can be chosen as the ratio of the Heisenberg
coupling parameters along the legs and rungs, while $t$ is the
coupling strength along the legs 
\cite{Poilblanc10,Cirac11,Peschel11,Lauchli11,Schliemann12}. 
Other recent work done in a similar spirit on coupled Luttinger liquids and
spin ladders includes Ref.~\cite{Lundgren13,Chen13}.
In the situation of quantum Hall bilayers at total filling factor of unity
\cite{Schliemann11} a natural choice for $\lambda$ is the layer separation 
in units of the magnetic length $\ell$, and $t$ is (a multiple of)
the Coulomb energy scale $e^2/ \ell$. Moreover, for fermionic hopping
models on lattices \cite{Peschel11,Schliemann13}, 
$\lambda$ is naturally chosen to be the ratio of
amplitudes for hopping within and between the subsystems, whereas $t$
is (proportional to) the former amplitude.

We note that Eq.~(\ref{enterg}) holds in prominent cases but not
truly in general. For an instructive counterexample see Ref.~\cite{Lundgren12}
where a spin ladder of clearly nonidentical legs was considered.
Besides, as the aforementioned studies show \cite{Poilblanc10,Cirac11,Peschel11,Lauchli11,Schliemann12,Schliemann11,Schliemann13}, a proportionality
of the type (\ref{enterg}) should in general not be expected to hold outside
the strong-coupling regime. Moreover, a critical assessment of the information 
in general included in an 
entanglement spectrum was given very recently in  Ref.~\cite{Chandran13}.
These authors define the entanglement Hamiltonian by
${\cal H}_{\rm ent}=-\ln\rho_{\rm red}$, implying $Z\equiv 1$, and argue
that the thermal state decsribed by $\rho_{\rm red}$ should be viewed
as being at an effective temperature $T=1/ \beta=1$.

In the light of the above developments we here further examine the concept of 
the entanglement
temperature and entanglement thermodynamics \cite{Schliemann11,Schliemann13}.
The leading question of this work is: ``When can a reduced density matrix of a 
many-body system be
considered as a thermal state, and, if so, what is its temperature?''
As an important point, the definition (\ref{rhoentham}) of the entanglement
Hamiltonian is not unique as one can always add any multiple of the unit 
operator to ${\cal H}_{\rm ent}$ 
which is compensated by the partition function $Z$. If this additional
term is independent of $\lambda$ (and therefore, as we shall see below in more
detail, independent of the entanglement temperature), such a change
has no significant effect. A nontrivial situation occurs if such
contributions depend on $\lambda$. The latter situation generically arises since
${\cal H}_{\rm ent}$ will in general contain all orders in that coupling
parameter. In the following section
\ref{genthermo} we review the formalism developed in Ref.~\cite{Schliemann13}
and discuss how to eliminate the above ambiguity. The latter issue does
essentially not occur for lattice models of free fermions 
\cite{Peschel11,Schliemann13,Peschel03} since the results given in Refs.
\cite{Peschel03,Cheong04} 
naturally lead to an entanglement Hamiltonian in terms of
non-interacting fermions without any  `artificial'  constant contribution.
The canonical thermodynamics arising from such Hamiltonians are described
in detail in section \ref{fthermo}. An important special case
occurs if the entanglement spectrum comprises only two levels
just differing in sign.
A nontrivial realization of this situation is, within mean-field
approximation, given by quantum Hall bilayers at total filling factor
$\nu=2$ \cite{Zheng97,DasSarma98,MacDonald99,Schliemann00}, 
as we discuss in section \ref{nu=2}. We close with a summary and an
outlook in section \ref{sumout}.

\section{Entanglement Thermodynamics}
\label{genthermo}

The entropy and the energy following from the reduced density matrix written
in the form (\ref{rhoentham}) read
\begin{eqnarray}
S & = & \langle-\ln\rho_{\rm red}\rangle\,,\\
\bar E & = & \langle {\cal H}_{\rm ent}\rangle
\end{eqnarray}
with $\langle\cdot\rangle={\rm tr}(\rho_{\rm red}\cdot)$. The bar over 
$\bar E$ indicates that this quantity should receive some refinement as
the derivative $\partial S/ \partial\bar E$ will in general fail to
be proportional to $\beta\propto\lambda$ at small $\lambda$, i.e. strong
coupling \cite{Schliemann13}. To define an energy $E$ fulfilling
the thermodynamic relation
\begin{equation}
\frac{\partial S}{\partial E}
=\frac{\partial S}{\partial \lambda}\frac{\partial \lambda}{\partial E}
=\beta(\lambda)
\label{dS/dE}
\end{equation}
we redefine the Hamiltonian as
\begin{equation}
{\cal H}_{\rm ent}(\lambda)
=\beta(\lambda){\cal H}_{\rm can}(\lambda)\,.
\label{redefham}
\end{equation} 
in terms of the {\em canonical entanglement Hamiltonian} 
${\cal H}_{\rm can}(\lambda)$.
The inverse thermodynamic temperature $\beta(\lambda)$
as a function of $\lambda$ is now determined via Eq.~(\ref{dS/dE}): 
Introducing the thermodynamic inner energy 
\begin{equation}
E(\lambda)=\langle{\cal H}_{\rm can}(\lambda)\rangle
\end{equation}
along with the free energy
\begin{eqnarray}
F(\lambda) & = & E(\lambda)-S(\lambda)/ \beta(\lambda)\\
 & = & -\ln(Z(\lambda))/ \beta(\lambda)
\end{eqnarray}
it is easy to see that Eq.~(\ref{dS/dE}) is equivalent to
\begin{equation}
\beta\frac{\partial\bar F}{\partial\beta}=\bar E
\label{Fbar}
\end{equation}
with $\bar F=\beta F=\bar E-S$. Thus, one has
\begin{equation}
\frac{\partial\ln\beta}{\partial\lambda}
=\frac{1}{\bar E}\frac{\partial\bar F}{\partial\lambda}
=\frac{1}{\bar E}\frac{\partial(\bar E-S)}{\partial\lambda}\,,
\label{betalambda1}
\end{equation}
which is the desired equation connecting the phenomenological
inverse temperature scale $\lambda$ to thermodynamic one  $\beta(\lambda)$. 
Indeed, a very similar relation is found in standard thermodynamics
between different temperature scales describing the same equilibrium 
states of a given body \cite{Landau80,Schwabl06}.

Let us now explore consequences of Eq.~(\ref{betalambda1}) related
to the ambiguity of constant contributions to the entanglement
Hamiltonian mentioned already in the introduction.
Expanding the the entanglement Hamiltonian in a power series in the coupling
parameter $\lambda$.
\begin{equation}
{\cal H}_{\rm ent}(\lambda)=\lambda h_1+\lambda^2h_2+\lambda^3h_3+\cdots\,,
\label{entpow}
\end{equation}
the leading order is, according to Eq.~(\ref{enterg}), determined by
the energy Hamiltonian of the subsystem,
\begin{equation}
h_1={\cal H}_1/ t\,.
\end{equation}
For the pertaining thermodynamic quantities one finds
\begin{eqnarray}
Z(\lambda) & = & {\rm tr}(\mathbf{1})
\Biggl(1-\lambda\frac{{\rm tr}\left(h_1\right)}{{\rm tr}(\mathbf{1})}
-\lambda^2\left(\frac{{\rm tr}\left(h_2\right)}{{\rm tr}(\mathbf{1})}
-\frac{1}{2}\frac{{\rm tr}\left(h_1^2\right)}{{\rm tr}(\mathbf{1})}\right)
\nonumber\\
& & \quad-\lambda^3\left(\frac{{\rm tr}\left(h_3\right)}{{\rm tr}(\mathbf{1})}
-\frac{{\rm tr}\left(h_1h_2\right)}{{\rm tr}(\mathbf{1})}
+\frac{1}{6}\frac{{\rm tr}\left(h_1^3\right)}{{\rm tr}(\mathbf{1})}\right)
\Biggr)+{\cal O}\left(\lambda^4\right)
\end{eqnarray}
and 
\begin{eqnarray}
F(\lambda) & = & -\ln{\rm tr}(\mathbf{1})
+\lambda\frac{{\rm tr}\left(h_1\right)}{{\rm tr}(\mathbf{1})}
+\lambda^2\left(\frac{{\rm tr}\left(h_2\right)}{{\rm tr}(\mathbf{1})}
-\frac{1}{2}\left(
\frac{{\rm tr}\left(h_1^2\right)}{{\rm tr}(\mathbf{1})}
-\left(\frac{{\rm tr}\left(h_1\right)}{{\rm tr}(\mathbf{1})}
\right)^2\right)\right)\nonumber\\
& & \quad+\lambda^3\Biggl(\frac{{\rm tr}\left(h_3\right)}{{\rm tr}(\mathbf{1})}
+\frac{1}{6}
\left(\frac{{\rm tr}\left(h_1^3\right)}{{\rm tr}(\mathbf{1})}
-3\frac{{\rm tr}\left(h_1\right){\rm tr}\left(h_1^2\right)}
{\left({\rm tr}(\mathbf{1})\right)^2}
+2\left(\frac{{\rm tr}\left(h_1\right)}{{\rm tr}(\mathbf{1})}
\right)^3\right)
\nonumber\\
&  & \qquad-\frac{{\rm tr}\left(h_1h_2\right)}{{\rm tr}(\mathbf{1})}
+\frac{{\rm tr}\left(h_1\right){\rm tr}\left(h_2\right)}
{\left({\rm tr}(\mathbf{1})\right)^2}\Biggr)+{\cal O}\left(\lambda^4\right)\,,
\end{eqnarray}
\begin{eqnarray}
\bar E(\lambda) & = & 
\lambda\frac{{\rm tr}\left(h_1\right)}{{\rm tr}(\mathbf{1})}
+\lambda^2\left(\frac{{\rm tr}\left(h_2\right)}{{\rm tr}(\mathbf{1})}
-\frac{{\rm tr}\left(h_1^2\right)}{{\rm tr}(\mathbf{1})}
+\left(\frac{{\rm tr}\left(h_1\right)}{{\rm tr}(\mathbf{1})}
\right)^2\right)\nonumber\\
& & \quad+\lambda^3\Biggl(\frac{{\rm tr}\left(h_3\right)}{{\rm tr}(\mathbf{1})}
+\frac{1}{2}\frac{{\rm tr}\left(h_1^3\right)}{{\rm tr}(\mathbf{1})}
-\frac{3}{2}\frac{{\rm tr}\left(h_1\right){\rm tr}\left(h_1^2\right)}
{\left({\rm tr}(\mathbf{1})\right)^2}
\nonumber\\
&  & \qquad-2\frac{{\rm tr}\left(h_1h_2\right)}{{\rm tr}(\mathbf{1})}
+2\frac{{\rm tr}\left(h_1\right){\rm tr}\left(h_2\right)}
{\left({\rm tr}(\mathbf{1})\right)^2}\Biggr)+{\cal O}\left(\lambda^4\right)\,.
\end{eqnarray}
In case ${\rm tr}\left(h_1\right)\neq 0$, the r.h.s of Eq.~(\ref{betalambda1})
reads
\begin{equation}
\frac{\partial\ln\beta}{\partial\lambda}
=\frac{1}{\lambda}+{\cal O}(1)\,,
\label{bpropl1}
\end{equation}
such that 
\begin{equation}
\beta(\lambda)=k\lambda+{\cal O}\left(\lambda^2\right)\,,
\label{bpropl2}
\end{equation}
where the  integration constant $k$ reflects a unit chosen to measure $\beta$
and has the same meaning as Boltzmann\rq s constant in standard thermodynamics.
Thus, as expected, the inverse temperature is at strong coupling proportional
to $\lambda$. However, if ${\rm tr}\left(h_1\right)=0$ Eq.~(\ref{bpropl1})
holds only if additionally ${\rm tr}\left(h_2\right)=0$ leading to
\begin{equation}
\frac{\partial\ln\beta}{\partial\lambda}
=\frac{1}{\lambda}
-2\frac{{\rm tr}\left(h_3\right)}{{\rm tr}\left(h_1^2\right)}
+\frac{{\rm tr}\left(h_1h_2\right)}{{\rm tr}\left(h_1^2\right)}
+{\cal O}\left(\lambda\right)\,.
\label{bpropl3}
\end{equation}
Thus, demanding
\begin{equation}
{\rm tr}\left({\cal H}_{\rm ent}(\lambda)-\lambda h_1\right)=0\,,
\label{tracecon1}
\end{equation}
i.e.
\begin{equation}
{\rm tr}\left(h_2\right)={\rm tr}\left(h_3\right)=\cdots=0\,,
\label{tracecon2}
\end{equation}
guarantees Eq.~(\ref{bpropl2}) and completely removes the ambiguity in the
definition of the entanglement Hamiltonian. In deriving Eq.~(\ref{bpropl3})
we have observed that ${\rm tr}(h^2_1)$ cannot be zero unless $h_1=0$, which,
by assumption, should not be the case.

To give a practical example, consider spin ladders described by the Hamiltonian
\begin{eqnarray}
{\cal H}=J_r\sum_i\vec S_{2i}\vec S_{2i+1}+J_l\sum_i\left(\vec S_{2i}\vec S_{2i+2}
+\vec S_{2i-1}\vec S_{2i+1}\right)
\end{eqnarray}
for spins of arbitrary but uniform length $S$ on $L$ rungs,
$i\in\{0,\dots,L-1\}$, with, for definiteness, periodic boundary conditions.
$J_r>0$ ($J_l>0$) is the antiferromagnetic
coupling along the rungs (legs). Defining $\lambda=2J_l/J_r$, the
low-order contributions to the entanglement Hamiltonian after tracing out
one leg are obtained via perturbation theory as
\cite{Schliemann12}
\begin{eqnarray}
h_1 & = & \sum_i\vec S_i\vec S_{i+1}\,,
\label{ladderh1}\\
h_2 & = & -\frac{1}{5}S(S+1)\sum_i\vec S_i\vec S_{i+2}\nonumber\\
 & & \quad-\frac{1}{20}\sum_i\left(
\left(\vec S_i\vec S_{i+1}\right)\left(\vec S_{i+1}\vec S_{i+2}\right)
+\left(\vec S_{i+1}\vec S_{i+2}\right)\left(\vec S_i\vec S_{i+1}\right)\right)
\label{ladders1}
\nonumber\\
 & & \quad-\frac{1}{12}\sum_i
\left(\left(\vec S_i\vec S_{i+1}\right)^2+2\vec S_i\vec S_{i+1}\right)
+\frac{1}{36}(S(S+1))^2L\,,
\label{ladderh2}
\end{eqnarray}
where $L$ is the number of rungs, and the constant term in 
Eq.~(\ref{ladderh2}) has been adjusted according to 
${\rm tr}\left(h_2\right)=0$. We note that for the smallest non-trivial
spin length $S=1/2$ the second-order term simplifies to
\cite{Lauchli11,Schliemann12}
\begin{equation}
h_2=-\frac{1}{8}\sum_i\left(\vec S_i\vec S_{i+2}-\vec S_i\vec S_{i+1}\right)\,,
\end{equation}
and no contribution proportional to the unit operator is required.
We note that the first-order result (\ref{ladders1}) holds in fact 
not just for spin ladders but also in various more general geometries
\cite{Schliemann12}; for a recent summary on spin systems and their 
entanglement spectra see also Ref.~\cite{Alet14}.

Now, using the condition ${\rm tr}(h_3)=0$ one finds
from Eq.~(\ref{bpropl3})
\begin{equation}
\frac{\partial\ln\beta}{\partial\lambda}
=\frac{1}{\lambda}-\frac{1}{8}+{\cal O}\left(\lambda\right)
\label{bpropl4}
\end{equation}
such that
\begin{equation}
\beta=k\lambda-\frac{k}{8}\lambda^2+{\cal O}\left(\lambda^3\right)\,.
\label{bpropl5}
\end{equation}
Remarkably, the second-order correction to the inverse temperature
(\ref{bpropl5}) is independent of the spin length $S$. However, we expect
such a dependence to occur in higher orders in $\lambda$.

\section{Free Fermions}
\label{fthermo}

Given the ground state $|\Psi(\lambda)\rangle$ of the full lattice system
of free fermions, consider the correlation matrix
\begin{equation}
{\cal C}_{\alpha\beta}(\lambda)
=\langle\Psi(\lambda)|b_{\alpha}^+b_{\beta}|\Psi(\lambda)\rangle
\end{equation}
where $b_{\alpha}^+$, $b_{\alpha}$ describe fermions in the remaining subsystem
using some arbitrary basis. The correlation determines the 
entanglement Hamiltonian via \cite{Peschel03,Cheong04}
\begin{equation}
{\cal H}_{\rm ent}(\lambda)=
\ln\left({\cal C}^{-1}(\lambda)-{\mathbf 1}\right)\,.
\end{equation}
More specifically, one has
\begin{equation}
{\cal H}_{\rm ent}=\sum_n\xi_na_n^+a_n\,,
\end{equation}
where the entanglement levels $\xi_n(\lambda)$ are given by
\begin{equation}
\xi_n(\lambda)=\ln\left(\frac{1-\eta_n(\lambda)}{\eta_n(\lambda)}\right)
\end{equation}
with $\eta_n$ being an eigenvalue of ${\cal C}$. 
The above fermionic operators read
\begin{equation}
a^+_n(\lambda)=\sum_{\alpha}U_{n \alpha}(\lambda)b_{\alpha}^+
\end{equation}
where the unitary matrix $U$ diagonalizes ${\cal C}$,
\begin{equation}
U(\lambda){\cal C}(\lambda)U^+(\lambda)
={\rm diag}\left(\eta_1(\lambda),\eta_2(\lambda),\dots\right)\,.
\end{equation}
We note that both the eigenvalues $\xi_n(\lambda)$ of the entanglement
Hamiltonian and its eigenstates created by the operators
$a^+_n(\lambda)$ are in general functions of $\lambda$.

The reduced density matrix can now be formulated as
\begin{equation}
\rho_{\rm red}(\lambda)=\frac{e^{-{\cal H}_{\rm ent}(\lambda)}}{Z(\lambda)}
=\prod_n\frac{e^{-\xi_n(\lambda)a_n^+a_n}}{1+e^{-\xi_n(\lambda)}}\,,
\end{equation}
along with the related quantities governing the thermodynamics,
\begin{eqnarray}
S(\lambda) & = & \sum_n\left(
\frac{\ln\left(1+e^{-\xi_n(\lambda)}\right)}{1+e^{-\xi_n(\lambda)}}
+\frac{\ln\left(1+e^{\xi_n(\lambda)}\right)}{1+e^{\xi_n(\lambda)}}
\right)\nonumber\\
& = & \sum_n\left(\ln\left(1+e^{-\xi_n(\lambda)}\right)
+\frac{\xi_n(\lambda)}{1+e^{\xi_n(\lambda)}}
\right)\,,\\
\bar E(\lambda) & = & \sum_n\frac{\xi_n(\lambda)}{1+e^{\xi_n(\lambda)}}\,,
\end{eqnarray}
such that
\begin{equation}
\bar F=\bar E-S=-\sum_n\ln\left(1+e^{-\xi_n}\right)
\end{equation}
and 
\begin{equation}
\frac{\partial}{\partial\lambda}\left(\bar E-S\right)=-\sum_n
\frac{\partial_{\lambda}\xi_n}{1+e^{\xi_n}}\,.
\end{equation}
A particularly simple but physically meaningful situation arises if 
the entanglement spectrum consists of just two levels differing in sign, 
$\xi_{\pm}=:\pm\xi$. Then one has
\begin{equation}
\bar E=-\xi\tanh\left(\frac{\xi}{2}\right)\qquad,\qquad
\frac{\partial\left(\bar E-S\right)}{\partial\lambda}
=-\left(\partial_{\lambda}\xi\right)\tanh\left(\frac{\xi}{2}\right)
\end{equation}
leading to 
\begin{equation}
\frac{\partial\ln\beta}{\partial\lambda}
=\frac{\partial_{\lambda}\xi}{\xi}
\label{betalambda2}
\end{equation}
with the very simple solution
\begin{equation}
\beta(\lambda)=k\xi(\lambda)\,,
\label{betalambda3}
\end{equation}
where the constant $k$ is the same as in Eq.~(\ref{bpropl2}).
Moreover, the spectrum of the canonical Hamiltonian is independent of 
$\lambda$ (or $\beta$), 
\begin{equation}
{\cal H}_{\rm can}=\frac{1}{k}\left(a_+^+a_+-a_-^+a_-\right)\,.
\label{betalambda4}
\end{equation}
The derivative of the entropy with respect to $\lambda$ is given by
\begin{equation}
\frac{\partial S}{\partial\lambda}=
\frac{-\xi\partial_{\lambda}\xi}{1+\cosh\xi}\,,
\end{equation}
such that the specific heat can be formulated as
($kT(\lambda)=1/ \xi(\lambda)$)
\begin{equation}
C=T\frac{\partial S}{\partial T}=\frac{\xi^2}{1+\cosh\xi}
=2\left(\frac{\xi/2}{\cosh\left(\xi/2\right)}\right)^2\geq 0\,.
\end{equation}
This quantity approaches zero for strong coupling, $\xi\to 0$, as well
as in the limit of vanishing coupling, $\xi\to\infty$. It attains a maximum
at $\xi=\xi_0$ determined by
\begin{equation}
\frac{\xi_0}{2}\tanh\left(\frac{\xi_0}{2}\right)=1\,,
\end{equation}
i.e. $\xi_0\approx 2.4$, such that
\begin{equation}
C\left(\xi_0\right)=\frac{2}{\sinh^2\left(\xi_0/2\right)}
=\frac{\xi_0^2}{2}-2\approx 0.88
\end{equation}
and
\begin{equation}
\bar E\left(\xi_0\right)=-2\,.
\end{equation}

\section{Quantum Hall Bilayers at $\nu=2$}
\label{nu=2}

In quantum Hall bilayers at total filling factor $\nu=2$, 
a canted antiferromagnetic phase
occurs as a function of the Zeeman gap and the tunneling amplitude.
Here the polarization directions of electron spins in both layers form
a nontrivial angle with each other. This phase separates a spin-polarized
phase and a spin-singlet phase. The theoretical prediction of this 
phase within Hartree-Fock approximation \cite{Zheng97,DasSarma98,MacDonald99}
was qualitatively (and, in part, also quantitatively)
confirmed by an exact-diagonalization study 
\cite{Schliemann00} and
has stimulated many further theoretical investigations; for most recent
works see, e.g, \cite{Hama12,Hama13}, older literature is summarized in
\cite{Ezawa08}. Also from the experimental point of view, the
existence of such a phase can by now be seen as established
\cite{Pellegrini97,Sawada98,Khrapai00,Fukuda06,Kumada07}.

Here we shall follow the formalism of Ref.~\cite{MacDonald99} where
a large part of the ground state structure was obtained within Hartree-Fock
theory in a fully analytical fashion. The Hartree-Fock approximation
reduces the Hamiltonian to a system of free fermions such that our above 
findings are easily applied. In particular, due to the projection onto the
lowest Landau level, the resulting entanglement spectrum will have just two
distinct levels.

\subsection{Hartree-Fock Theory}

The total Hamiltonian of the system is the sum of the Coulomb interaction
projected onto the lowest Landau level and the 
single-particle Hamiltonian
\begin{equation}
h=-\frac{\Delta_v}{2}\tau^z-\frac{\Delta_t}{2}\tau^x-\frac{\Delta_z}{2}\sigma^x
\label{biham}
\end{equation}
where the Pauli matrices $\vec\tau$ and $\vec\sigma$ act on the 
layer (pseudo-)spin and genuine electron spin, respectively.
$\Delta_v$ is a bias voltage between the quantum wells, and $\Delta_t$
describes the tunneling between them. $\Delta_z$ is the Zeeman energy
due to the perpendicular magnetic field which is chosen here (following
Ref.~\cite{MacDonald99}) to lie along the $x$-direction in electron spin space.
Orbital states in the lowest Landau level are labeled by
a quantum number $X$ which depends on the chosen gauge. Neglecting
charge fluctuations within the layers, each mode $X$ is filled with two
electrons leading to the following single Slater determinant
as a variational ansatz for the ground state,
\begin{equation}
|\Psi[z]\rangle=\prod_X
\sum_{k,l=1}^4\left(z^1_kz^2_lb^+_k(X)b^+_l(X)\right)|0\rangle\,.
\end{equation} 
Here $z^1$, $z^2$ are two orthonormal spinors,
where $k=1,2$ describe a spin-up(down) electron in the upper layer,
while $k=3,4$ refer to a spin-up(down) electron in the lower layer.
Ref.~\cite{MacDonald99} uses the parameterization
\begin{eqnarray}
z^1 & = & \left(
\cos\frac{\vartheta}{2}\cos\frac{\chi_{\uparrow}}{2},
\sin\frac{\vartheta}{2}\sin\frac{\chi_{\downarrow}}{2},
\cos\frac{\vartheta}{2}\sin\frac{\chi_{\uparrow}}{2},
-\sin\frac{\vartheta}{2}\cos\frac{\chi_{\downarrow}}{2}
\right)\,,\\
z^2 & = & \left(
-\sin\frac{\vartheta}{2}\sin\frac{\chi_{\uparrow}}{2},
\cos\frac{\vartheta}{2}\cos\frac{\chi_{\downarrow}}{2},
\sin\frac{\vartheta}{2}\cos\frac{\chi_{\uparrow}}{2},
\cos\frac{\vartheta}{2}\sin\frac{\chi_{\downarrow}}{2}
\right)
\end{eqnarray}
in terms of three angles $\vartheta$, $\chi_{\uparrow}$, and $\chi_{\downarrow}$,
which turns out to be flexible enough to describe the Hartree-Fock ground state
as a function of various system parameters.
Defining $\chi_{\pm}=(\chi_{\downarrow}\pm\chi_{\uparrow})/2$, the variational energy
per mode is
\begin{eqnarray}
\varepsilon & = & -\Delta_v\cos\vartheta\cos\chi_+\cos\chi_-
-\Delta_t\cos\vartheta\sin\chi_+\cos\chi_-
-\Delta_z\sin\vartheta\sin\chi_-\nonumber\\
& & -F_+-2H\cos^2\vartheta\cos^2\chi_+\cos^2\chi_-\nonumber\\
& & -F_-\Bigl(\sin^2\vartheta\cos^2\chi_+\sin^2\chi_-
+\cos^2\vartheta\cos^2\chi_+\cos^2\chi_-\nonumber\\
& & \qquad\qquad+\sin^2\chi_+\sin^2\chi_-\Bigr)
\label{varerg}
\end{eqnarray}
with
\begin{eqnarray}
F_{\pm} & = & \frac{1}{(2\pi)^2}\int d^2qe^{-q^2\ell^2/2}V_{\pm}(\vec q)\,,\\
H & = & \frac{1}{2\pi\ell^2}V_-\left(\vec q=0\right)\,.
\end{eqnarray}
Here $\ell$ is the magnetic length, and $V_{\pm}(\vec q)$ describe the
Coulomb interaction within and between the layers. Assuming quantum wells
of negligible width, one has (using obvious notation)
\begin{equation}
V_{\pm}(\vec r)=\frac{e^2}{2\epsilon}
\left(\frac{1}{r}\pm\frac{1}{\sqrt{r^2+d^2}}\right)
\end{equation}
where $d$ is the layer separation.

\begin{figure}
 \includegraphics[width=\columnwidth]{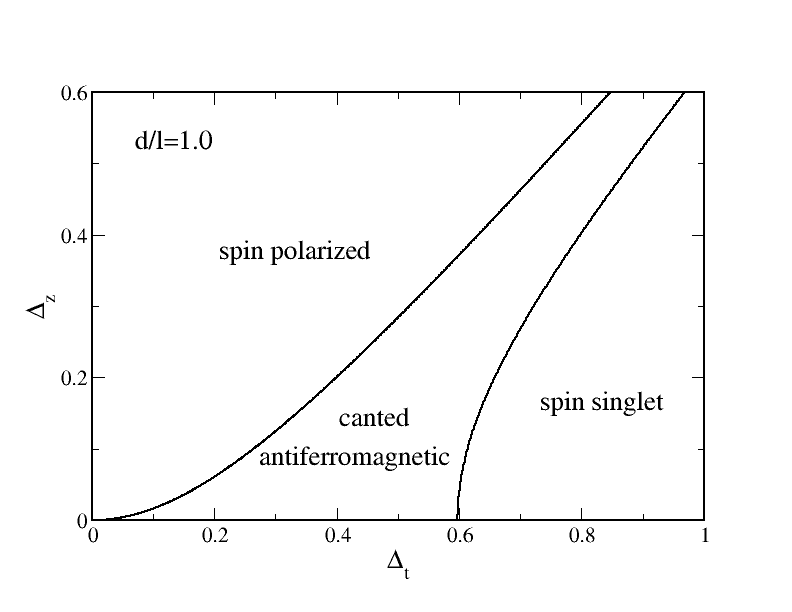}
\caption{Phase diagram at zero bias voltage and layer separation $d=\ell$
as a function of Zeeman energy $\Delta_z$ and tunneling gap $\Delta_t$. The
canted antiferromagnetic phase separates the spin-polarized phase 
and the spin-singlet phase. All energies are given in units of the 
Coulomb energy scale $e^2/(\epsilon\ell)$.}
\label{fig1}
\end{figure}

For unbiased layers, $\Delta_v=0$, minimization of (\ref{varerg}) leads
to $\chi_+=\pi/2$ and
\begin{eqnarray}
\sin^2\vartheta & = & 
-\frac{\Delta_z^2\left(\left(\Delta_t^2-\Delta_z^2\right)^2
-\left(2\Delta_tF_-\right)^2\right)}
{\left(\Delta_t^2-\Delta_z^2\right)^3}\,,
\label{mincon1}\\
\sin^2\chi_- & = & 
\frac{\left(2\Delta_tF_-\right)^2-\left(\Delta_t^2-\Delta_z^2\right)^2}
{4F_-^2\left(\Delta_t^2-\Delta_z^2\right)}\,,
\label{mincon2}
\end{eqnarray}
fulfilling
\begin{equation}
\tan\vartheta=\frac{\Delta_z}{\Delta_t}\tan\chi_-\,,
\label{mincon3}
\end{equation}
provided that the above r.h.s of Eqs.~(\ref{mincon1}), (\ref{mincon2})
are nonnegative and do not exceed unity.
This is the case if $\Delta_z^{\rm min}\leq\Delta_z\leq\Delta_z^{\rm max}$
with
\begin{eqnarray}
\Delta_z^{\rm min}& = & \sqrt{\Delta_t(\Delta_t-2F_-)}\,,
\label{delmin}\\
\Delta_z^{\rm max} & = & \sqrt{F_-^2+\Delta_t^2}-F_-\,.
\label{delmax}
\end{eqnarray}
For $\Delta_z^{\rm min}<\Delta_z<\Delta_z^{\rm max}$ the bilayer system is in the
{\em canted antiferromagnetic phase} where the order parameter
\begin{equation}
\langle\tau^z\sigma^z\rangle=2\cos\vartheta\sin\chi_+\sin\chi_-
\end{equation}
is nonzero. This phase separates the {\em spin-polarized phase} 
($\Delta_z>\Delta_z^{\rm max}$, $\vartheta=\chi_-=\pi/2$) and the 
{\em spin singlet phase}
($\Delta_z<\Delta_z^{\rm min}$, $\vartheta=\chi_-=0$). 
A phase diagram for typical
system parameters is given in Fig.~\ref{fig1}.
In the approximation-free exact-diagonalization study \cite{Schliemann00}
the phase boundary between the spin-polarized phase and the
canted antiferromagnetic phase was found to be {\em exactly} given by
the Hartree-Fock result (\ref{delmax}), whereas quantitative corrections 
occur to the lower phase boundary (\ref{delmin}). As a result,
the canted antiferromagnetic phase turns out to be smaller than predicted
by mean-field theory, but definitely existing \cite{Schliemann00}.

\subsection{Entanglement Spectrum}

Tracing out, say, the top layer from the Hartree-Fock ground state, 
the entanglement spectrum is encoded in the correlation matrix 
${\cal C}_{kl}=\langle\Psi|b^+_kb_l|\Psi\rangle$, $k,l\in\{3,4\}$,
\begin{eqnarray}
{\cal C} & = & 
\frac{1}{2}\left(1-\cos\vartheta\cos\chi_+\cos\chi_-\right){\mathbf 1}_{2\times 2}
\nonumber\\
& & -\frac{1}{2}\sin\chi_-
\left(
\begin{array}{cc}
\cos\vartheta\sin\chi_+ & -\sin\vartheta \\
-\sin\vartheta & -\cos\vartheta\sin\chi_+
\end{array}
\right)\,.
\end{eqnarray}
At vanishing bias voltage ($\chi_+=\pi/2$) its eigenvalues
\begin{equation}
\eta_{\pm}=\frac{1}{2}\left(1\mp\sin\chi_-\right)
\end{equation}
are independent of $\vartheta$ and lead to the entanglement levels
\begin{equation}
\xi_{\pm} =\ln\frac{1-\eta_{\pm}}{\eta_{\pm}}
=\ln\frac{1\pm\sin\chi_-}{1\mp\sin\chi_-}
=\pm 2\,{\rm artanh}\left(\lambda/2\right)\,,
\end{equation}
where we have introduced the dimensionless
coupling parameter $\lambda=2\sin\chi_-$.
The entanglement Hamiltonian can be formulated as ($\xi=|\xi_{\pm}|$)
\begin{equation}
{\cal H}_{\rm ent}=\xi\sum_X\left(a^+_+(X)a_+(X)-a^+_-(X)a_-(X)\right)
\end{equation}
with
\begin{eqnarray}
a^+_+(X) & = & \cos\frac{\vartheta}{2}b^+_3(X)
-\sin\frac{\vartheta}{2}b^+_4(X)\,,\\
a^+_-(X) & = & \sin\frac{\vartheta}{2}b^+_3(X)
+\cos\frac{\vartheta}{2}b^+_4(X)\,,
\end{eqnarray}
where, as defined above, $b^+_3(X)$ ($b^+_4(X)$) creates an electron with
spin up (down) in the remaining bottom layer. At strong coupling
($\lambda\ll 1$) the entanglement Hamiltonian is given by
(cf. Eq.~(\ref{mincon3}))
\begin{equation}
{\cal H}_{\rm ent}=\lambda
\sum_X\left(b^+_3(X)b_3(X)-b^+_4(X)b_4(X)\right)
+{\cal O}\left(\lambda^2\right)\,.
\end{equation}
Thus, we have a relation of the type
(\ref{enterg}) with the energetic subsystem Hamiltonian
\begin{equation}
{\cal H}_1=\frac{\tilde\Delta_z}{2}
\sum_X\left(b^+_3(X)b_3(X)-b^+_4(X)b_4(X)\right)
\label{ham1}
\end{equation}
and $t=\tilde\Delta_z/2$ describing the exchange-enhanced Zeeman splitting in 
a quantum Hall monolayer at filling factor $\nu=1$. In the latter system we
have again neglected charge fluctuations, consistent with the above mean-field
approximation. Interestingly the effective field 
in the Hamiltonian (\ref{ham1}) points along the $z$-axis of spin space 
while the magnetic field in the original bilayer Hamiltonian (\ref{biham})
was in $x$-direction.

In summary, we have demonstrated that the mean-field theory of 
bilayer quantum Hall systems at total filling factor $\nu=2$
constitutes a system of free fermions with an entanglement spectrum
comprising only two different levels of the same modulus.
As seen in Eq.(\ref{betalambda4}), in such a case the
spectrum of the redefined canonical entanglement Hamiltonian is
independent of $\lambda$ (and therefore independent of $\beta$).
The redefined canonical entanglement Hamiltonian reads
according to Eq.~(\ref{betalambda4})
\begin{eqnarray}
{\cal H}_{\rm can} & = & 
\frac{1}{k}\sum_X\left(a^+_+(X)a_+(X)-a^+_-(X)a_-(X)\right)\\
& = & \frac{1}{k}\sum_X\left[
\left(b_3^+(X),b_4^+(X)\right)
\left(
\begin{array}{cc}
\cos\vartheta & -\sin\vartheta \\
-\sin\vartheta & -\cos\vartheta
\end{array}
\right)
\left(
\begin{array}{c}
b_3(X)\\
b_4(X)
\end{array}
\right)\right]\,.
\label{hamcan}
\end{eqnarray}
Moreover, $\vartheta$ and $\chi_-$ are determined via 
Eqs.~(\ref{mincon1}), (\ref{mincon2}) by $\Delta_z$, $\Delta_t$
which are independent parameters {\em within} the canted antiferromagnetic
phase but bound to each other at the phase boundaries
(\ref{delmin}), (\ref{delmax}). Likewise, $\vartheta$ and $\chi_-$ are
independent parameters within the canted antiferromagnetic
phase outside the phase boundaries where they are identical.
In this sense, the canonical entanglement Hamiltonian (\ref{hamcan})
is independent of $\lambda=2\sin\chi_-$ and therefore independent of the
inverse temperature given in Eq.~(\ref{betalambda3}).

\section{Summary and Outlook}
\label{sumout}

We have extended previous studies on the relationship between the 
entanglement spectrum of a composite many-body system and the
energy spectrum of a subsystem. Inspired by the recent literature, 
the basic question investigated here is under which circumstances
a reduced density matrix of a many-body system allows an interpretation 
as a thermal
state, and, if so, how to determine its temperature.
For strong coupling between the subsystems, the entanglement Hamiltonian is,
in a variety of important cases, proportional to the energy Hamiltonian
of the subsystem with the proportionality factor $\lambda$
being an appropriately defined
coupling parameter. It is suggestive to interpret this quantity
as an inverse temperature $\beta(\lambda)$. 
Indeed, for the regime away from strong coupling,
a differential equation (\ref{betalambda1}) for $\beta(\lambda)$ 
ensuring the fulfillment of standard thermodynamic relations has been given
\cite{Schliemann13}. This construction is based on the redefinition
(\ref{redefham}) of the entanglement Hamiltonian. We note, however, that
the redefined Hamiltonian ${\cal H}_{\rm can}(\lambda)$ will in general
depend on the coupling parameter $\lambda$ and, in turn, on $\beta(\lambda)$,
as illustrated in the present work on the example of spin ladders.

On the other hand, the definition of a Hamiltonian generating a given
density matrix contains generally an ambiguity as one
can always add a multiple of the unit operator to the former without
changing the latter.
Demanding that all contributions to the entanglement Hamiltonian of higher than
linear order in the coupling parameter should have vanishing trace removes
this ambiguity and guarantees the above interpretation as an inverse
temperature to hold. In particular, this condition
includes the case where the entanglement Hamiltonian is in total
traceless. This case is realized, e.g., for spin ladders of 
arbitrary spin length, and for many fermionic hopping models on lattices.
A more detailed analysis of the latter type of systems shows that
the analogy to standard thermodynamics is closest if the entanglement spectrum
consists of just two levels differing in sign. In this case the spectrum of the
redefined entanglement Hamiltonian ${\cal H}_{\rm can}$
is independent of $\lambda$. A nontrivial example for such a
situation is provided by the mean-field theory of 
bilayer quantum Hall systems at total filling factor $\nu=2$. As an additional
feature, here also the 
eigenstates (and not only the eigenvectors) of ${\cal H}_{\rm can}$ are
independent of temperature.

Obviously, the results obtained here 
call for further study. For instance, it would be
interesting to possibly identify other situations where the spectrum
of ${\cal H}_{\rm can}$ (or ${\cal H}_{\rm can}$ in total) is independent of 
the entanglement temperature. Indeed, it is the fact that 
${\cal H}_{\rm ent}(\lambda)$
in general fails to be linear in $(\lambda)$ which makes the 
{\em entanglement thermodynamics} developed here different from standard
thermodynamics.
On the other hand, a Hamiltonian depending on a
temperature is not that unfamiliar to theoretical physics since many effective
(Hartree-Fock type) descriptions of many-body systems involve, e.g., 
temperature-dependent occupation numbers. 

Moreover, a deeper physical understanding of the condition (\ref{tracecon1})
eliminating the abovementioned ambiguity in the definition of the entanglement
Hamiltonian is desirable. This goal might be achieved by applying and 
testing the formalism developed here on other physical systems.

\section*{Acknowledgements}

This work was supported by DFG via SFB631.

{}

\end{document}